\documentclass[12pt]{article}

\RequirePackage[numbers]{natbib}
\usepackage{array}
 \usepackage{amsmath,amssymb}
 \usepackage{amsthm}
 \usepackage{bm}
 \usepackage{latexsym}
 \usepackage{CJK}
 \usepackage{multirow}
 \usepackage{caption}
 \usepackage{graphicx,color}
 \usepackage{amssymb,amsfonts}
 \usepackage{hyperref}
 \usepackage[english]{babel}
 \usepackage{times}
 \usepackage{enumerate}
\usepackage{geometry}
\usepackage{float}
\usepackage{booktabs}
\usepackage{endnotes}
\usepackage{mathtools}
\usepackage{amscd}
\usepackage{bbm}
\usepackage{multirow}
\usepackage{booktabs}
\usepackage{graphicx}
\usepackage{url}
\usepackage{hyperref}
\usepackage{setspace}
\usepackage{adjustbox}
\usepackage{fixltx2e}
\usepackage[dvipsnames]{xcolor}

\usepackage[scaled]{beramono}
\usepackage[T1]{fontenc}
\usepackage{amsmath,amsfonts,amssymb}
\usepackage{mathtools}
\usepackage{url}
\usepackage{hyperref}
\usepackage{setspace}

\usepackage{fixltx2e}
\usepackage[dvipsnames]{xcolor}

\usepackage[toc]{appendix}

\usepackage[utf8]{inputenc}
\usepackage{authblk}
\usepackage{subcaption}

\newtheorem{innercustomgeneric}{\customgenericname}
\providecommand{\customgenericname}{}
\newcommand{\newcustomtheorem}[2]{%
  \newenvironment{#1}[1]
  {%
   \renewcommand\customgenericname{#2}%
   \renewcommand\theinnercustomgeneric{##1}%
   \innercustomgeneric
  }
  {\endinnercustomgeneric}
}

\newcustomtheorem{customlemma}{lemma}

\usepackage{authblk}

\DeclareMathOperator*{\argmin}{argmin}

\begin{document}

\def\spacingset#1{\renewcommand{\baselinestretch}%
{#1}\small\normalsize} \spacingset{1}

		\title{Optimal Transport for Latent Integration with An Application to Heterogeneous Neuronal Activity Data}

	\author[1]{Yubai Yuan}
	\author[2]{Babak Shahbaba}
	\author[3]{Norbert Fortin}
	\author[3]{Keiland Cooper}
	\author[4]{Qing Nie}
	\author[2]{Annie Qu \footnote{Corresponding author: aqu2@uci.edu}}
	\affil[1]{Department of Statistics, The Pennsylvania State University}
	\affil[2]{Department of Statistics, University of California Irvine}
	\affil[3]{Department of Neurobiology and Behavior, University of California Irvine}
	\affil[4]{Department of Mathematics, Department of Developmental and Cell Biology, Department of Biomedical Engineering, University of California Irvine}


		\date{ }
		\maketitle
		\vspace{-8mm}

\spacingset{1.76}		

		\begin{abstract}

{Detecting dynamic patterns of task-specific responses shared across heterogeneous datasets is an essential and challenging problem in many scientific applications in medical science and neuroscience. 
In our motivating example of rodent electrophysiological data, identifying the dynamical patterns in neuronal activity associated with ongoing cognitive demands and behavior is key to uncovering the neural mechanisms of memory. One of the greatest challenges in investigating a cross-subject biological process is that the systematic heterogeneity across individuals could significantly undermine the power of existing machine learning methods to identify the underlying biological dynamics. In addition, many technically challenging neurobiological experiments are conducted on only a handful of subjects where rich longitudinal data are available for each subject. The low sample sizes of such experiments could further reduce the power to detect common dynamic patterns among subjects. In this paper, we propose a novel heterogeneous data integration framework based on optimal transport to extract shared patterns in complex biological processes. The key advantages of the proposed method are that it can increase discriminating power in identifying common patterns by reducing heterogeneity unrelated to the signal by aligning the extracted latent spatiotemporal information across subjects. Our approach is effective even with a small number of subjects, and does not require auxiliary matching information for the alignment. In particular, our method can align longitudinal data across heterogeneous subjects in a common latent space to capture the dynamics of shared patterns while utilizing temporal dependency within subjects. Our numerical studies on both simulation settings and neuronal activity data indicate that the proposed data integration approach improves prediction accuracy compared to existing machine learning methods.} 

\end{abstract}

\noindent\textbf{Key words:} Distribution alignment, Electrophysiological study, Gromov-Wasserstein barycenter, Latent embedding, Optimal transport, Semi-supervised learning


\section{Introduction}

New emerging and rapidly expanding technologies for conducting high throughput scientific studies have allowed the simultaneous recording of activity from large ensembles of neurons during complex behavior \citep{allen2013evolution, eichenbaum2014time}. These technical advances are expected to lead to paradigm-shifting breakthroughs in our fundamental understanding of information processing in the brain, as well as to a novel platform to investigate sources of deficits and potential treatments in neurological disorders. However, at this time, the main limiting factor toward achieving this goal is that existing statistical and computational methods are insufficient to address the new challenges associated with  extracting and integrating the essential  information contained in these large and complex multidimensional datasets. Among these challenges, by far the most difficult task is how to properly integrate data across multiple subjects from a highly heterogeneous population \citep{hackam2006translation, perel2007comparison}. This paper provides a fundamental solution and a rigorous framework to address this issue.

Our proposed method is motivated by, and applied to, a recent electrophysiological neuroscience study,   
which was designed to elucidate the neuronal mechanisms underlying memory for sequences of events, an ability essential to our normal daily life function. Although it is well established that this ability depends on the integrity of the hippocampus, a region near the center of our brain, it remains unclear \textit{how} the activity of ensembles of hippocampal neurons can trigger forms of memory. Since the structure and function of the hippocampus is well observed across mammals, this type of information is primarily obtained by recording neural activity in the hippocampus of rodents due to the invasiveness of the approach. 
In particular, considerable rodent research has shown that hippocampal neurons can encode sequences of locations, but there has been no direct evidence that the coding property can be extended to the non-spatial memories which are typically investigated in human studies.

\begin{figure}[h]
\centering
\includegraphics[scale=0.38]{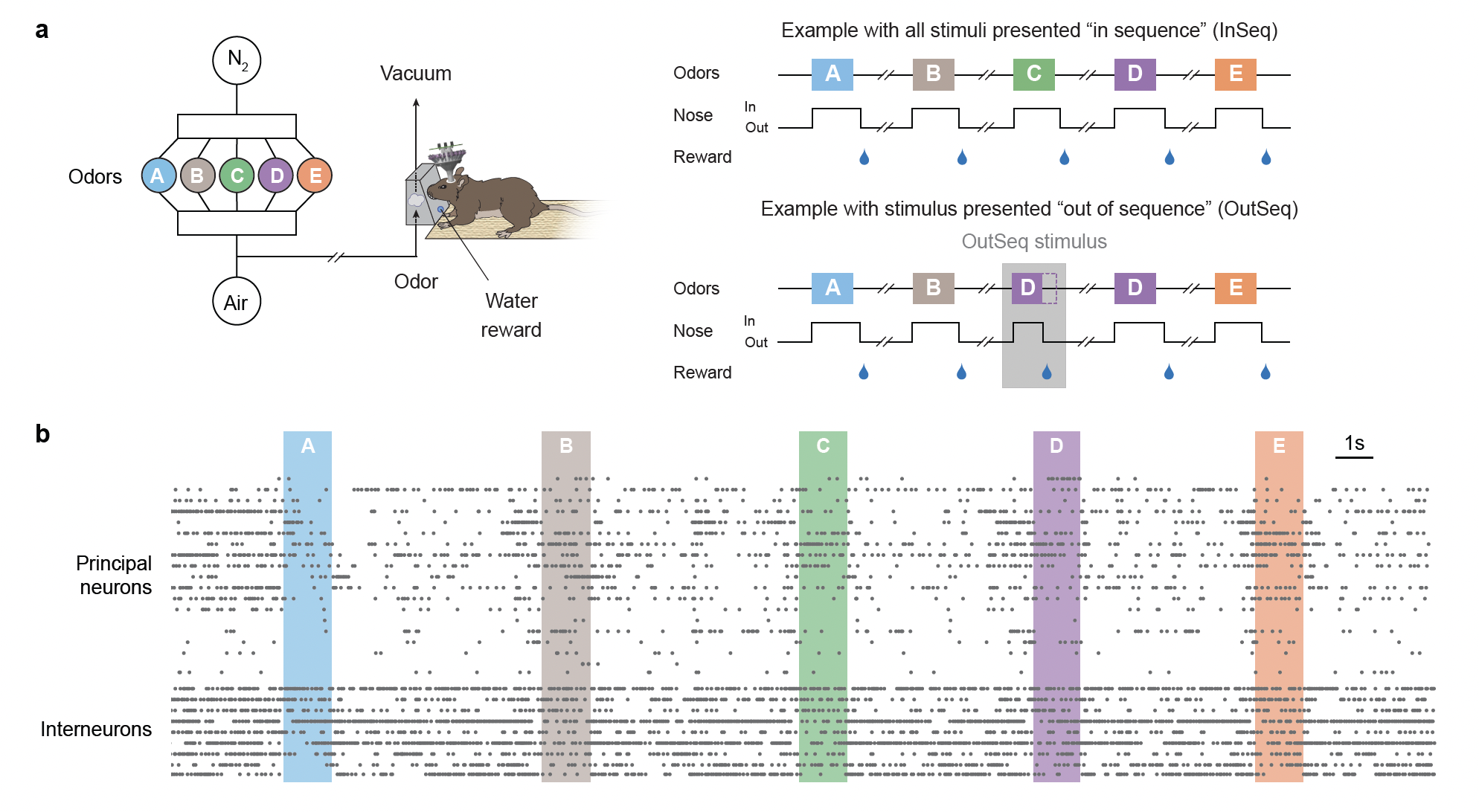}
\renewcommand{\baselinestretch}{1}
\caption{{Neural activity was recorded from hippocampal region CA1 as animals performed a complex non-spatial sequence memory task. (\textbf{a}) The task involves repeated presentations of sequences of non-spatial events (odor stimuli) to subjects (rodents). Using an automated odor delivery system (top left), sequences of five odors were presented in the same odor port. (\textbf{b)} Example ensemble activity from representative subject during one sequence presentation.}} 
\label{fig:experiment}
\end{figure}

The  experiment recorded neural activity in the CA1 region of the hippocampus as rats performed a nonspatial sequential memory task such as  testing a  sequence of 5 odors (see Figure \ref{fig:experiment}) where neuron functional activity was recorded by neuronal electrophysiological data \citep{shahbaba2022hippocampal}. 
Although  statistical machine learning methods can be used to extract and visualize latent representations of sequential neuronal  
activities associated with different tasks in each subject, 
it is still a great challenge to integrate subject-specific  latent representations across different subjects. Yet this data integration is a crucial step, as the primary objective of such studies is to identify fundamental coding properties which are common across subjects \citep{akil2011challenges, sejnowski2014putting}.


To integrate cross-subject neuronal data, there are several neuronal data integration methods designed to directly extract the cross-subject latent spaces, i.e., model the subject-wise neuronal signal as the reconstruction from a set of reference neuronal dynamics, which are estimated in the common latent space via Gaussian process \citep{ebrahimi2020time}, canonical correlation analysis \citep{zhang2017inter}, and principal component analysis \citep{linden2022movement}. 
These methods assume that the subject-wise signals are linear transformations of the common latent dynamics, which implies that the heterogeneity among subjects can be captured via linear transformations between their neuronal data. This assumption can be restrictive and oversimplified when the individual heterogeneity in neuronal activities is complex and strong.

In this paper, we propose to aggregate information by 
finding the common relations between neuron activity patterns and experimental outcomes 
among different subjects. On the other hand, this type of  data  is known to contain noise, redundancies, and subject-specific idiosyncrasies, which poses great difficulties in developing methods to identify the underlying common neural patterns among subjects. 
One of the most significant challenges is the intrinsic heterogeneity within neuron activities \citep{churchland2007temporal, marinelli2014heterogeneity}. Specifically, the underlying mapping between neuron activity pattern and a specific outcome can be differentially expressed among different subjects due to systematic differences across subjects caused by random deviation from common patterns. Also, this kind of experiment is typically conducted on only a handful of subjects where high-dimensional neuron signal data are available from each subject \citep{churchland2007temporal, marinelli2014heterogeneity}. The systematic heterogeneity across individuals and low sample size significantly undermines the capacity of traditional machine learning  methods to identify cross-subjects patterns. To alleviate this problem, a new data integration method is highly needed to properly account for heterogeneity in order to discern the underlying common patterns across subjects.   
Specifically, we propose a novel integrated latent alignment (ILA) framework by incorporating deep representation learning \citep{bengio2009learning, hinton2006reducing, bengio2013representation} and optimal transport (OT) \citep{villani2009optimal, agueh2011barycenters} to integrate heterogeneous data.

{Compared with existing data integration methods, optimal transport can relax the linearity assumption to reconstruct the nonlinear transformation between individual spike sequences and cross-subject spike patterns, and therefore one can integrate heterogeneous spike data with more flexible distributions from subjectwise perturbations.
In addition, the optimal-transport-based barycenter is able to integrate multiple distributions of neuronal encodings to obtain shared geometric structures \citep{peyre2019computational}, which enables one to detect common clustering patterns from heterogeneous neural encodings of different subjects' spike data, and preserve common spike patterns via filtering out subjects' heterogeneity when aligning spike data.}

The proposed ILA framework consists of three stages: First, we use deep learning to project the temporal neural data to a low-dimensional latent space for each subject, which allows us to identify the macro neuron patterns 
from each subject. 
Notice that the representations of the neuron activity are projected into different latent spaces for different subjects, or the compressed neural features from each subject. 
Therefore, in the second stage, we develop a novel data integration method based on optimal transport \citep{agueh2011barycenters, courty2014domain, titouan2019optimal}. Specifically, we incorporate cross-subject heterogeneity by aligning both the geometric structure of the latent neural features and the feature-response patterns from different subjects. The cross-subject alignment allows us to aggregate subject-wise data in a shared latent space. In the last stage, we adopt either supervised learning  or unsupervised learning methods in the aligned latent space to capture the common neural patterns across subjects. 

The proposed ILA  has several significant advantages. First, it provides a principle strategy to integrate data from heterogeneous subjects and increase statistical power in identifying the common patterns. 
In addition, the integration is effective even when the number of subjects is small. Secondly, the proposed method does not require restrictive assumptions on the mappings for the latent feature spaces from heterogeneous subjects. Instead, the ILA utilizes the smoothness in association  between neural activity features and  responses to align the patterns directly, which allows for perturbations between responses and features across subjects. Finally, the proposed method can incorporate heterogeneous temporal latent features, and task-specific machine learning algorithms. Both the numerical studies on simulations and the electrophysiological data show that the proposed data integration method improves the performance of classification tasks compared with methods using individual data only. 

This paper is organized as follows: Section 2 introduces the background of the electrophysiological study and optimal transport. Section 3 introduces the proposed data integration method for supervised learning and the corresponding optimization algorithms. Section 4 demonstrates simulation studies, and Section 5 illustrates an application of the proposed method to an electrophysiological study. The last section provides conclusions and discussion.

\section{Background and Notations}

In this section, we provide a brief introduction to our motivating data from the electrophysiological study, and then introduce the optimal transport techniques related to our method. The electrophysiological data were collected as rats performed a non-spatial sequence memory task \cite{allen2014sequence}. Briefly, the task involves repeated presentations of a sequence of odors (odors A,B,C,D, and E) in the same odor port, and requires subjects to determine whether each odor is presented “in sequence” (InSeq; e.g., AB\underline{C}...) or “out of sequence” (OutSeq; e.g., AB\underline{D}...) to receive a water reward (Figure \ref{fig:experiment}a). Note that the task was self-paced in that each odor presentation was initiated by the rat placing its nose in the port (median interval between consecutive odors $\sim$5 s). In each session, the same sequence was presented multiple times (Figure \ref{fig:experiment}a, \textit{right}), with approximately half the presentations including all stimuli InSeq (Figure \ref{fig:experiment}a,\textit{top)} and the other half including one stimulus OutSeq (Figure \ref{fig:experiment}a,\textit{bottom}). Incorrect InSeq/OutSeq judgments resulted in termination of the sequence. Neural spiking activity was recorded using 22 tetrodes (bundles of four electrodes) in rats tested on the well-trained sequence (ABCDE). For each odor presentation (trial), the data typically features spike counts from $\sim$50-90 neurons, and trial identifiers (e.g., odor presented, InSeq/OutSeq, response correct/incorrect). Figure \ref{fig:experiment}b illustrates examples of ensemble activity (putative principal neurons and interneurons) from a representative subject during one sequence presentation. More detailed description of the data is in the Section 11 of Supplemental Materials.

In summary, this dataset is well-suited to develop and test the required methodology to integrate neural activity data across subjects. First, the dataset was shown to contain complex and distributed forms of coding that were undetectable when examining the activity of individual neurons independently, but emerged when considering the activity of each ensemble of neurons as a whole. Second, the size of the dataset is in a tractable range (50-90 neurons per subject); the ensembles are large enough to detect emergent coding properties, but small enough to allow the iterative development and testing of analytical and computational solutions. Third, the task structure provides points of convergence across subjects (e.g., common sequences of stimuli and associated neural states) as well as points of divergence (e.g., errors by a specific animal).


Here, we use {optimal transport (OT), which has been widely applied in machine learning fields (e.g., domain adaptation and transfer learning), 
to provide a principled way of aligning distributions while preserving the geometric structures. The classical OT based on the Wasserstein distance requires that data distributions be defined on the same space, which may not be satisfied in many real applications 
where data could arise from different domains of distributions. To overcome this limitation, the Gromov–Wasserstein distance \citep{memoli2014gromov} and fused Gromov–Wasserstein distance \citep{vayer2020fused} are introduced to extend the concept of OT to cross-domain distribution alignment.

\noindent \textbf{Gromov-Wasserstein distance}$\;$  Consider two datasets $\{(X^{(s)},Y^{(s)}), P^{(s)}\}\; (s = 1,2)$ from two different metric-measure spaces where $(X^{(s)}, Y^{(s)})$ denotes a set of the covariates and the responses for the $s$th dataset. Notice that $X^{(1)}$ and  $X^{(2)}$ can be from different domains with different data types and dimensions of features. The $P^{(s)}$ denotes the probability mass distribution on the samples from the $s$th dataset to represent  relative importance for the corresponding samples. If no prior information is known on the $P^{(s)}$, one can set the empirical distribution version of $P^{(s)} = \frac{1}{N_s}\mathbbm{1}^{\top}_{N_s}$ with mass of each sample point being $\frac{1}{N_s}$, where $\mathbbm{1}_{N_s} = (1,1,\cdots,1)_{N_s}$ and $N_s$ is the sample size of the $s$th dataset. To quantify the discrepancy between distributions on different domain spaces, \citep{memoli2014gromov} proposes the Gromov-Wasserstein distance on $\{X^{(s)},P^{(s)}\}, (s = 1,2)$ to measure the geometric structural similarity of two distributions. Specifically, one constructs the distance matrix $D^{(s)} = (d^{(s)}_{ij}), (s=1,2)$ to represent similarities among samples within the $s$th dataset where $d^{(s)}_{ij} = d^{(s)}(X^{(s)}_i,X^{(s)}_j)$ is the distance between $X^{(s)}_i$ and $X^{(s)}_j$ corresponding to a given metric $\bm{X}\times \bm{X} \rightarrow R^{+}$ in the $s$th space. The Gromov-Wasserstein distance is defined as
\vspace{-5mm}
$$\text{GW}(D^{(1)},D^{(2)},P^{(1)}, P^{(2)}) = \min_{\Omega \in \bm{\Omega}_{\{P^{(1)}, P^{(2)}\}}} \sum_{i, j, k, l} L\left(D^{(1)}_{i, k}, D^{(2)}_{j, \ell}\right)\cdot \Omega_{i, j}\cdot \Omega_{k, \ell},$$
where $D^{(1)}_{i,k}$ and $D^{(2)}
{j,l}$ denote the $(i,k)$ element and $(j,l)$ element in $D^{(1)}$ and $D^{(2)}$, and $\Omega = (\Omega_{ij})$ is a transport coupling matrix between two datasets satisfying
$$\Omega \in \bm{\Omega}_{\{P^{(1)}, P^{(2)}\}} = \left\{\Omega \in\left(\mathbb{R}_{+}\right)^{N_{1} \times N_{2}} : \Omega \mathbbm{1}^{\top}_{N_{2}}=P^{(1)}, \Omega^{\top} \mathbbm{1}^{\top}_{N_{1}}=P^{(2)} \right\}.$$
Here $L$ is a loss function to account for the element-wise discrepancy between the distance matrices. Typical choices of $L$ include the quadratic loss $L(a; b) = (a - b)^2$ and Kullback-Leibler divergence $L(a; b) = a \log(a/b) - a + b$. 

\noindent \textbf{Fused Gromov-Wasserstein distance} To further measure the discrepancy of joint distribution $(X^{(s)},Y^{(s)})$ between two datasets, the fused Gromov-Wasserstein distance is a generalization of the aforementioned Gromov-Wasserstein distance. To incorporate the response information $Y^{(s)}$ into the alignment between two datasets, we can first construct $M = \big(d(Y^{(1)}_i,Y^{(2)}_j)\big)_{ij}$ which is a $N_1\times N_2$ distance matrix, and $d(\cdot,\cdot)$ stands for a similarity measurement between responses within the same metric space. Then the fused Gromov-Wasserstein distance is defined as:
\vspace{-4mm}
{\begin{align*}
\text{FGW}(D^{(1)},D^{(2)},M,P^{(1)}, P^{(2)}) = \min_{\Omega \in \bm{\Omega}_{\{P^{(1)}, P^{(2)}\}}} \sum_{i, j, k, l} \big\{ (1-\alpha) M_{ij} + \alpha L\left(D^{(1)}_{i, k}, D^{(2)}_{j, \ell}\right)\big\}\!\cdot\! \Omega_{i, j}\!\cdot\! \Omega_{k, \ell},
\end{align*}} 
where $0\leq \alpha \leq 1$. The FGW distance searches for a coupling $\Omega$ between two datasets that minimizes a weighted cost combining the mass transportation on response from $Y^{(1)}_i$ to $Y^{(2)}_j$, and on covariate pairs from $D^{(1)}_{i,k}$ to $D^{(2)}_{j,l}$. 
The $i$th column in the coupling matrix $\Omega$ decides the proportions of probability mass in the $i$th sample point from the first dataset to be transported to each sample point in the second dataset. Notice that when $\alpha = 1$, the FGW distance degenerates to the GW distance. 

\textbf{Optimal transport for data integration} The optimal transport can be applied to cross-dataset alignment. We consider two different 
covariate spaces $(X^{(1)}, d^{{(1)}}, \mu^{(1)})$ and $(X^{(2)}, d^{{(2)}}, \mu^{(2)})$ where $d^{(1)},d^{(2)}$ and $\mu^{(1)},\mu^{(2)}$ denote the metrics and probability measures on each covariate space. 
The optimal transport can be used to align the geometric structures of $\mu^{(1)}$ and $\mu^{(2)}$. Specifically, we randomly sample pairs of covariates $\{(x_i,x_j)\}$ from ${X}^{(1)} \times {X}^{(1)}$ and pairs of covariates $\{(y_k,y_l)\}$ from ${X}^{(2)} \times {X}^{(2)}$, respectively. Given that the pairwise metrics $\{d^{(1)}(x_i,x_j)\}$ and $\{d^{(2)}(y_k,y_l)\}$ capture the corresponding geometric information of $\mu^{(1)}$ and   $\mu^{(2)}$, we aim to seek an alignment plan between $\mu^{(1)}$ and $\mu^{(2)}$ to minimize the discrepancy between the two metrics $L((x_i,x_j),(y_k,y_l)) = | d^{(1)}(x_i,x_j) - d^{(2)}(y_k,y_l)|$ at the population level. 
Based on the set of joint distributions $\bm{\Omega} = \{\Omega \in P({X}^{(1)} \times {X}^{(2)}) \mid \Omega({X}^{(1)} \times A ) =  \mu^{(2)}(A),\Omega(B \times {X}^{(2)} ) =  \mu^{(1)}(B)\}$, the optimal alignment plan to minimize geometric structure discrepancy between $\mu^{(1)}$ and $\mu^{(2)}$ is defined as:
\begin{align}\label{eq(1)}
\Omega =\underset{\Omega \in \bm{\Omega}}{\operatorname{argmin}} \int_{ ( {X}^{(1)} \times {X}^{(2)} ) \times ( {X}^{(1)} \times {X}^{(2)} ) } \!\!\!\!\!\!L((x_i,x_j),(y_k,y_l)) d \Omega\left(\mathbf{x}_i, \mathbf{y}_k\right)d \Omega\left(\mathbf{x}_j, \mathbf{y}_l\right).
\end{align}
{Notice that the metric discrepancy $L((x_i,x_j),(y_k,y_l))$ is defined on the product space, therefore we use the product of alignment plan $\Omega\left({x}_i, {y}_k\right)\times \Omega\left({x}_j, {y}_l\right)$ to serve as the alignment between sample pairs $(x_i,x_j)$ and $(y_k,y_l)$. The alignment plan obtained in (\ref{eq(1)}) is well-defined and induces the Gromov–Wasserstein distance which generalizes the Wasserstein distance. We extend the above intuition into the proposed methods in Section 3.}

{
\section{Methodology}

In this section, we propose a new data integration framework ILA for both supervised and unsupervised learning tasks. The proposed ILA framework consists of three stages: data compression, cross-subject alignment, and common pattern learning, which is illustrated in Figure \ref{ILA}. {The data compression in stage 1 of the ILA framework directly adopts existing dimension reduction methods such as principal component analysis \citep{linden2022movement} or the autoencoder \citep{kingma2019introduction}. Autoencoders have been used to compress neural activity data in other applications \citep{wu2018deep,speiser2017fast, burbank2015mirrored, shahbaba2022hippocampal} and empirical studies have demonstrated their superior performance in capturing non-linear features from high-dimensional data compared with other unsupervised dimensionality reduction methods such as PCA \citep{almotiri2017comparison}, Isomap, and locally linear embedding \citep{wang2016auto}. Therefore, we utilize autoencoder in stage 1, and focus on the key ingredients stage 2 and stage 3 of our method that utilizes the idea of incorporating the heterogeneity domains in datasets by aligning them into a common data space. We first introduce a cross-domain data integration method for the supervised learning task in Section 3.1, and then introduce an integration method for the cross-domain temporal patterns in Section 3.2.} 
\vspace{-3mm}
\begin{figure}[h]
\centering
\includegraphics[scale=0.80]{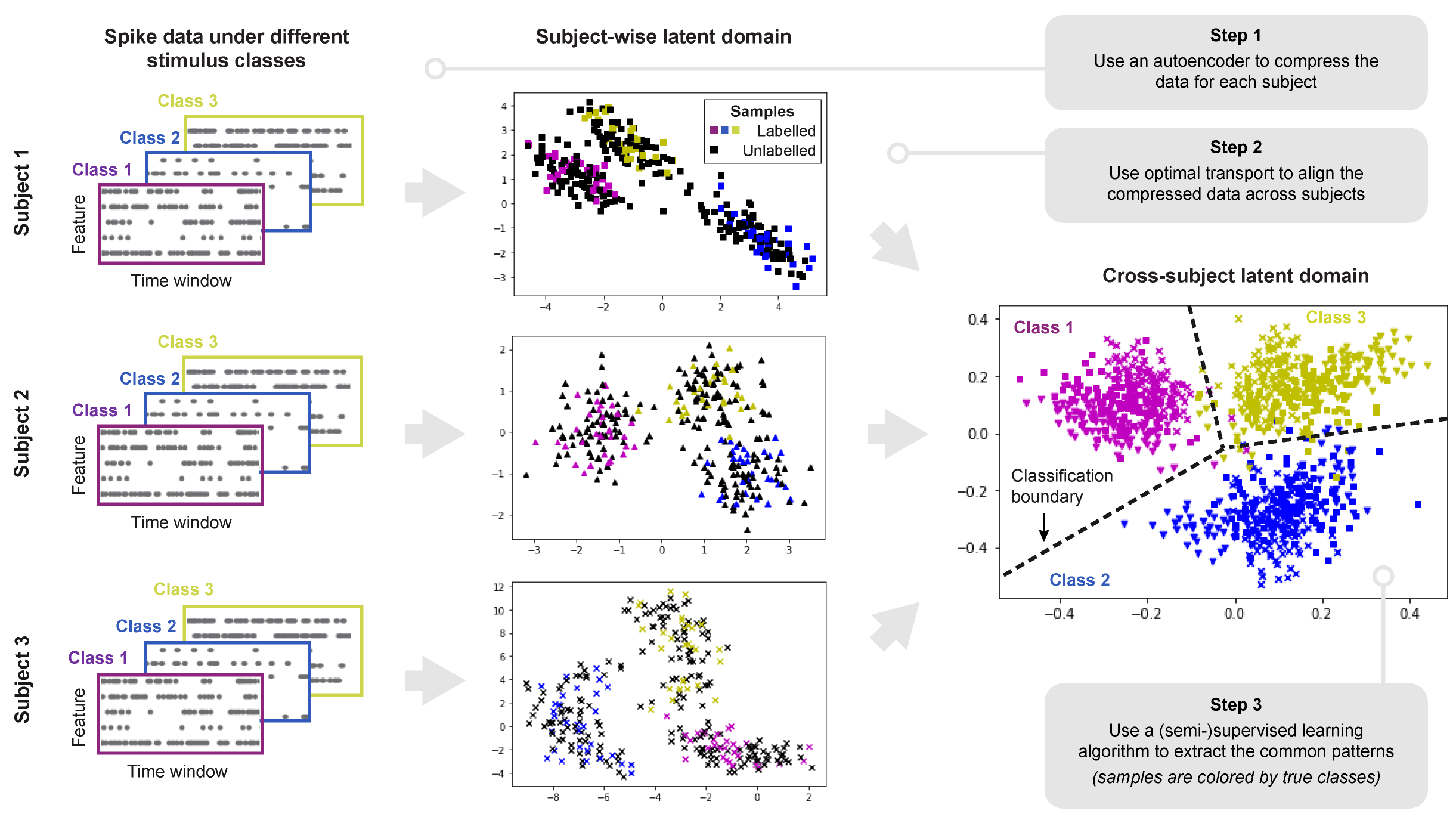}
\renewcommand{\baselinestretch}{1}
\caption{{The proposed integrated latent alignment framework (ILA). In stage 1, we utilize an autoencoder to compress the electrophysiological data for each subject. In stage 2, we use optimal transport to align the compressed data from different subjects in latent spaces. In the last stage, a supervised learning algorithm is implemented in the aligned latent space to extract common patterns.}} 
\label{ILA}
\end{figure}
\vspace{-3mm}


\subsection{Integrated supervised learning on multiple heterogeneous datasets}

In this subsection, we consider multiple datasets $\{ (X^{(s)}, Y^{(s)})\}_{s = 1}^S$ defined in $S$ different domains. The joint distributions $\mathbb{P}^{(s)}(\bm{X}^{(s)},\bm{Y}^{(s)})$ on the $s$th dataset can be distinct from one another due to the heterogeneity among marginal distributions $\mathbb{P}^{(s)}(\bm{X}^{(s)}),\; s = 1,2,\cdots, S$. To enable the cross-domain alignment, we assume that the marginal distributions $\mathbb{P}^{(s)}(\bm{X}^{(s)}),\; s = 1,2,\cdots, S$ share the same geometric structure in that for two datasets $s$ and $s'$: there exists a probabilistic mapping $\Gamma^{s',s}: \bm{X}^{(s')} \times \bm{Y}^{(s')} \to \bm{X}^{(s)}\times  \bm{Y}^{(s)} $  satisfying
\begin{align}\label{OT_assump}
\mathbb{P}^{(s)}\big(\bm{X}^{(s)},\bm{Y}^{(s)}\big) \approx \mathbb{P}^{(s')}\big(\Gamma^{s',s}(\bm{X}^{(s')},\bm{Y}^{(s')})\big), 1\leq s \neq s' \leq S. 
\end{align}  
Notice that the measure-preserving mapping $\Gamma^{s',s}$ corresponds to the alignment plan $\Omega$ in (\ref{eq(1)}) between two spaces $\bm{X}^{(s')} \times \bm{Y}^{(s')}$ and $\bm{X}^{(s)}\times  \bm{Y}^{(s)}$. In addition, the geometry information of $\mathbb{P}^{(s)}(\bm{X}^{(s)})$ can be represented as a set of pairwise distances $\{d^{(s)}(x_i,x_j)\}$ where $x_i, x_j$ are samples from $\mathbb{P}^{(s)}(\bm{X}^{(s)})$. Then the distribution of $\{d^{(s)}(x_i,x_j)\}$  represents the concentration of samples from $\bm{X}^{(s)}$ at different regions. Therefore, a common geometric structure between $\bm{X}^{(s)}$ and $\bm{X}^{(s')}$ implies that the mapping $\Gamma^{s',s}$ preserves distances between the two domains as
\begin{align*}
d^{(s')}\big(x_i, x_j\big) \approx  d^{(s)}\big(\Gamma^{s',s}(x_i), \Gamma^{s,s'}(x_j)\big), 1\leq s \neq s' \leq S.
\end{align*}   
The cross-domain relation (\ref{OT_assump}) guarantees the existence of a shared distribution of response conditional on predictors across $S$ datasets as follows:
\begin{align*}
\mathbb{P}^{(s)}\big(\bm{Y}^{(s)}|\bm{X}^{(s)}\big) \approx \mathbb{P}^{(s')}\big(\bm{Y}^{(s')}|\Gamma^{s',s}(\bm{X}^{(s')})\big), 1\leq s \neq s' \leq S. 
\end{align*}
This enables us to increase the effective training sample size in the $s$ dataset via integrating labeled data from other datasets $s'$ associated with $s$, whereby the performance of the machine learning model of the $s$th domain can be improved. 

Without loss of generality, we consider the classification task as follows. For the $s$th dataset $1 \leq s \leq S$, let the training dataset and the set of unlabeled sample points be denoted as $(X_{l}^{(s)},Y_{l}^{(s)})$ and $X_{p}^{(s)}$, respectively. We denote the sample points in the training set and unlabeled dataset as $(X_{l,i}^{(s)},Y_{l,i}^{(s)})$ and $X_{p,i}^{(s)}$ such that $X_{l}^{(s)} = \{X_{l,i}^{(s)}\}_{i=1}^{n_l^{(s)}}, Y_{l}^{(s)} = \{Y_{l,i}^{(s)}\}_{i=1}^{n^{(s)}_l}$ and $X_{p}^{(s)} = \{X_{p,i}^{(s)}\}_{i=1}^{n^{(s)}_p}$, where $n^{(s)}_l$ and $n^{(s)}_p$ are the number of training samples and unlabeled samples in the $s$th dataset, respectively. In the following, we consider the data integration for the classification task on the $s$th dataset. The data integration on the other datasets follows the same procedure. The proposed method is summarized as follows:

\noindent \textbf{Step 1: Individual supervised learning} Given a specific classifier $f(\cdot)$, we train the classifier on $(X_{l}^{(s)},Y_{l}^{(s)})$ separately as $f_s(\cdot)$, and then predict on the unlabeled set $X^{(s)}_{p}$ to obtain predicted labels $\hat{Y}^{(s)}_{p} = f_s(X^{(s)}_{p})$ for each dataset $s = 1,\cdots, S$.

\noindent \textbf{Step 2: One-to-many cross-domain alignment} We consecutively align training sets from other $S-1$ datasets $(X_{l}^{(s')},Y_{l}^{(s')}),\; s' \neq s, 1\leq s'\leq S$ to the region of unlabeled data $X^{(s)}_{p}$ in the $s$th dataset. Specifically, we adopt the fused Gromov-Wasserstein optimal transport as follows: 
\begin{align}\label{FGW}
\Omega^{s,s'}  = \argmin_{\bm{\Omega}} \sum_{i, j, k, l} \big\{ (1-\alpha) M^{s,s'}_{ij} + \alpha L(D^{(s)}_{i, k}, D^{(s')}_{j, \ell})\big\} \Omega_{i, j} \Omega_{k, \ell}, \\
\bm{\Omega} = \left\{\Omega \in\left(\mathbb{R}_{+}\right)^{n_p^{(s)} \times n_l^{(s')} }: \Omega \mathbbm{1}_{n_l^{(s')}}= \frac{1}{n_p^{(s)}}\mathbbm{1}_{n_p^{(s)}}, \Omega^{\top} \mathbbm{1}_{n_p^{(s)}}= \frac{1}{n_l^{(s')}}\mathbbm{1}_{n_l^{(s')}} \right\}, \nonumber
\end{align}
where $M^{s,s'} = \big(d(\hat{Y}^{(s)}_{p,i},{Y}_{l,j}^{(s')})\big)$, $D^{(s)} = \big(d^{(s)}(X^{(s)}_{p,i},X^{(s)}_{p,j})\big)$, and $D^{(s')} = \big(d^{(s')}(X^{(s')}_{l,i},X^{(s')}_{l,j})\big)$ are distance matrices, $d(\cdot)$ is the similarity measure of the labels, and $d^{(s)}(\cdot)$ denotes the distance between predictors on the $s$th domain. The $i$th column in the coupling matrix $\Omega^{s,s'}$ decides the proportions of probability mass in the training sample point $(X^{(s')}_{l,i}, Y^{(s')}_{l,i})$ to be transported to each unlabeled sample point in $\{X^{(s)}_{p,i}\}$. 
We choose L$_2$ norm $L(\cdot) = \| \cdot \|^2$ in (\ref{FGW}).

Given that $X_l^{(s')}$ and $X_p^{(s)}$ are not in the same domain, $\Omega^{s,s'}$ intends to match the internal geometric structures between $X_l^{(s)}$ and $X_p^{(s')}$, which are measured by distance matrices $D^{(s)}$ and $D^{(s')}$, respectively. On the other hand, $\Omega^{s,s'}$ also aims to align the label distribution between datasets $s$ and $s'$ though the term $(1-\alpha)M^{s,s'}_{ij}$. Minimizing (\ref{FGW}) allows us to match the joint distribution $\mathbb{P}(X_l^{(s')}, Y_l^{(s')})$ to $\mathbb{P}(X_p^{(s)}, Y_p^{(s)})$ so that (\ref{OT_assump}) can be held where $\Omega^{s,s'}$ serves as the underlying cross-domain transformation $\Gamma^{s,s'}$, and hence we can integrate training data from heterogeneous domains to capture the common label-predictor pattern. Because $X_p^{(s)}$ are not labeled, i.e., the corresponding $Y_p^{(s)}$ are not observed, we approximate     
$\mathbb{P}(X_p^{(s)}, Y_p^{(s)})$ by $\mathbb{P}(X_p^{(s)}, \hat{Y}_p^{(s)})$
where, $\hat{Y}_p^{(s)}$ are obtained in Step 1.





\noindent \textbf{Step 3: Training data mapping} We map the training set $(X_{l}^{(s')},Y_{l}^{(s')})\; (s'\neq s)$ from other $S-1$ datasets to the domain of the $s$th dataset. Specifically, we reconstruct the $j$th training sample in the $s'$th dataset $(X_{l,j}^{(s')},Y_{l,j}^{(s')})$ on the $s$th domain as $(\hat{X}_{l,j}^{(s')},Y_{l,j}^{(s')})$, where 
\begin{align}\label{OT_sample}
\hat{X}_{l,j}^{(s')} = \underset{X}{\operatorname{argmin}}\sum_{i} \Omega^{s,s'}_{ij} d^{(s)}\left(X,X^{(s)}_{p,i}  \right),   
\end{align}  
and $\Omega^{s,s'}$ is obtained from Step 2. The barycentric mapping (\ref{OT_sample}) corresponds to
a weighted average of unlabeled data $X^{(s)}_p$ in the $s$th domain. Intuitively, the barycentric mapping reconstructs the training sample from the $s$the dataset in the domain of the $s'$th dataset. We denote the reconstructed samples as $\hat{X}^{(s')}_l =\{\hat{X}^{(s')}_{l,j}\}$. Based on Step 2 and assumption (\ref{OT_assump}), $(\hat{X}^{(s')}_l, Y^{(s')}_l)$ follows the same distribution in the  $s$ domain as $(X^{(s)}_p, Y^{(s)}_p)$.   

\noindent \textbf{Step 4: Retrain the model with augmented training data} We obtain reconstructed training data from other $S-1$ datasets as $(\tilde{X}^{(s)}_l,\tilde{Y}^{(s)}_l) = \underset{s'\neq s}{\cup}(\hat{X}^{(s')}_l, Y^{(s')}_l)$, and train the classifier $f(\cdot)$ on the augmented training dataset $(X^{(s)}_l \cup \tilde{X}^{(s)}_l,  Y^{(s)}_l \cup \tilde{Y}^{(s)}_l)$ as $f_s^{aug}(\cdot)$. Then we update the prediction on unlabeled data $X^{(s)}_p$ as $\hat{Y}^{(s)}_p = f_s^{aug}(X^{(s)}_p)$.  

{In summary, the proposed data integration method is designed for the semi-supervised learning scenario, where the training data from each dataset consists of a small proportion of labeled samples $\{X^{(s)}_{l}, Y^{(s)}_{l}\}_{s=1}^S$ and a large number of unlabeled samples $\{X^{(s)}_{p}\}_{s=1}^S$. The objective is to predict outcomes for the unlabeled samples.}

{
\subsection{Temporal integration of multiple heterogeneous dataset}  

In this subsection, we introduce the proposed data integration for supervised learning on multiple longitudinal heterogeneous datasets. Specifically, we consider $S$ longitudinal heterogeneous datasets $\{\big(X^{(s)}(t),Y^{(s)}(t)\big)\}_{s = 1}^S$ measured at time points $t = 1,\cdots,T$, where covariates $X^{(s)}(t)$, $s = 1,\cdots,S$ change over time, but are defined on different domains.

To perform data integration, we denote the temporal alignment mappings $\Gamma^{s',s}$ between domains $s'$ and $s$ as:
\begin{align}\label{OT_assump_2}
\mathbb{P}^{(s)}\big(\bm{X}^{(s)}(t),\bm{Y}^{(s)}(t)\big) \approx \mathbb{P}^{(s')}\Big(\Gamma^{s',s}\big(\bm{X}^{(s')}(t),\bm{Y}^{(s')}(t)\big)\Big), 1\leq s \neq s' \leq S, \; 1\leq t \leq T.
\end{align}
We consider the case where the alignment mappings $\{\Gamma^{s',s}\}$ are time invariant, and hence (\ref{OT_assump_2}) implies that the common temporal pattern is identifiable among the $S$ different datasets, i.e., 
\begin{align}\label{OT_assump_3}
\mathbb{P}^{(s)}\big(\bm{X}^{(s)}(t),\bm{Y}^{(s)}(t)\big) \approx \mathbb{P}\Big(\Gamma^{s}\big(\bm{Z}(t),\bm{Y}(t)\big)\Big), 1\leq s \leq S,\; 1\leq t \leq T,  
\end{align}
where $\{\bm{Z}(t)\}_{t=1}^T$ and $\{\bm{Y}(t)\}_{t=1}^T$ denote the covariate-associated and response-associated dynamic process on the aligned data domains, respectively. In other words, the common temporal pattern on the conditional distribution of response given that covariates is captured by $\{(\bm{Z}(t),\bm{Y}(t)\}_{t=1}^T$. Then the $s$th mapping $\{\Gamma^{s}\}_{s=1}^S$ projects the common temporal patterns to domains of the $s$th dataset. By inferring $\{(\bm{Z}(t),\bm{Y}(t)\}_{t=1}^T$ and $\{\Gamma^{s}\}_{s=1}^S$, we can integrate cross-domain temporal data into a common latent space to improve the performance of the supervised learning task as the effective training sample size increases. 

In the following, we denote the labeled training dataset as $\big(X^{(s)}_l(t), Y^{(s)}_l(t)\big)$ and the unlabeled dataset as $X^{(s)}_p(t)$, such that $X^{(s)}_l(t) = \{X^{(s)}_{l}(t)_i\}_{i=1}^{n_l^{(s)}}$, $Y^{(s)}_l(t) = \{Y^{(s)}_{l}(t)_i\}_{i=1}^{n_l^{(s)}} )$ and $X^{(s)}_p(t) = \{X^{(s)}_{p}(t)_i\}_{i=1}^{n_p^{(s)}}$, $1\leq s \leq S, 1\leq t\leq T$. The number of total samples in the $s$th dataset is $n^{(s)} = n_l^{(s)} + n_p^{(s)}$.  Let $X^{(s)}(t) = (X_l^{(s)}(t), X_p^{(s)}(t))$ be the set of covariates in the $s$th domain. The proposed method is summarized as follows:  

\noindent \textbf{Step 1: Individual supervised learning} Given a specific classifier $f(\cdot)$, we train the classifier on $(X_{l}^{(s)}(t),Y_{l}^{(s)}(t))$ separately as $f^{(s)}_t(\cdot)$ for each dataset $s$ and time point $t$, $1\leq s \leq S, 1\leq t \leq T$. Then obtain predicted labels through unlabeled data $X^{(s)}_{p}(t)$ via $\hat{Y}^{(s)}_{p}(t) = f_t^{(s)}\big(X^{(s)}_{p}(t)\big)$ for each dataset and time point. 

\noindent \textbf{Step 2: Cross-domain temporal alignment} We align the time-evolving datasets from $S$ different domains via the proposed temporal Fused Gromov-Wasserstains (FGW) barycenter. Denoting the augmented labels for the $s$th dataset as $ Y^{(s)}(t) = \big(Y^{(s)}_l(t),\hat{Y}^{(s)}_p(t)\big),1\leq t \leq T$, we estimate the alignment couplings $\{\Omega^{(s)}\}_{s=1}^S$ and the common dynamic process $\{C(t)\}_{t=1}^T = \{D(t),Y(t)\}_{t=1}^T$ in the following aligned space:
\begin{align}\label{temporal_alignment}
\{C(t)\}_{t=1}^T,& \{\Omega^{(s)}\}_{s=1}^S = \argmin_{\substack{C(t), t= 1,\cdots, T \\ \Omega^{(s)}\in \bm{\Omega}^{(s)}, s = 1,\cdots, S}} \sum_{t=1}^T \sum_{s=1}^S w_s\; \text{FGW}\Big( C(t),\Omega^{(s)}\Big),\\
\text{where FGW}\Big(C(t),\Omega^{(s)}\Big) & \!=\! \alpha\! \sum_{ij}\big(M^{(s)}(t)\big)_{ij}\Omega^{(s)}_{ij}\! +\! (1\!-\!\alpha)\!\sum_{i,j,u,v}\! \|\big(D^{(s)}(t)\big)_{ij}\! -\! \big(D(t)\big)_{uv}\|^2 \Omega^{(s)}_{iu}  \Omega^{(s)}_{jv},  \nonumber \\
\bm{\Omega}^{(s)} &= \left\{\Omega \in\left(\mathbb{R}_{+}\right)^{n^{(s)} \times n }: \Omega \mathbbm{1}_{n} = \frac{1}{n^{(s)}}\mathbbm{1}_{n^{(s)}}, \Omega^{\top} \mathbbm{1}_{n^{(s)}}= \frac{1}{n}\mathbbm{1}_{n} \right\}, \nonumber
\end{align}    
where $M^{(s)}(t) = \Big(d\big(Y^{(s)}(t)_i,Y(t)_j\big)\Big)$, $D^{(s)}(t) = \Big(d^{(s)}\big(X^{(s)}(t)_i,\; X^{(s)}(t)_j\big)\Big),\; 1\leq t \leq T$, $\alpha\in [0,1]$ is the trade-off parameter, and $n$ is the number of template samples in the common latent space. The similarity matrix $D(t) \in R_{+}^{n \times n}$ represents the pairwise distances aligned with $D^{(s)}(t)$ from different datasets at time $t$, and consists of the common covariates dynamics $Z(t)$ in (\ref{OT_assump_3}). Specifically, given the positions of $n$ template samples $\{Z(t)_i \}_{i=1}^n$ in the aligned latent space at time $t$, we have $D(t)_{ij} = \tilde{d}\big(Z(t)_i,Z(t)_j\big)$, where $\tilde{d}(\cdot)$ is the distance between $Z(t)_i$ and $Z(t)$
in the latent space. In addition, the transport coupling $\{\Omega^{(s)} \in [0,1)^{(n_l^{(s)}+n_p^{(s)})\times n}\}$ serves as the domain transformation $\{\Gamma^{s}\}$ in (\ref{OT_assump_3}) which assigns the probability mass on $\{X^{(s)}(t)_i, Y^{(s)}(t)_i\}_{i=1}^{n_l^{(s)}+n_p^{(s)}}$ for template samples $\{Z(t)_i, Y(t)_i\}_{i=1}^n$ in the latent space with $1\leq t \leq T$. Finally, $w_s\geq 0 $ in (\ref{temporal_alignment}) denotes the weights of the $s$th dataset in the alignment as $\sum_{s=1}^S w_s = 1$, and we set  $w_s = \frac{1}{S}$ by default. 
 
\noindent \textbf{Step 3: Data integration on aligned space} We embed the aligned similarity matrix $D(t)$ to a $k$-dimensional latent space at each time point $t$. Specifically, we infer $Z(t) = \{Z(t)_i \in R^k\}_{i=1}^n$ via
\begin{align*}
Z(t) = \argmin_{\bm{Z}}\sum_{i,j}^{n}( D(t)_{ij} - \| \bm{Z}_{i}-\bm{Z}_{j}\|_2 )^2.
\end{align*}  
Then we map all $s$ datasets to latent space $\{Z(t)\}_{t = 1}^T$, i.e., the training sample $\big(X^{(s)}_{l}(t)_i,Y^{(s)}_{l}(t)_i\big)$ and the $j$th testing sample $X^{(s)}_{p}(t)_j$ in the $s$th dataset are reconstructed in the latent space as $\big(\hat{X}^{(s)}_{l}(t)_i,Y^{(s)}_{l}(t)_i\big)$ and $\hat{X}^{(s)}_{p}(t)_j$ respectively, where
\begin{align*}
\hat{X}_{l}^{(s)}(t)_i = \underset{x}{\operatorname{argmin}}\sum_{u} \Omega^{s}_{iu} d\big(x,Z(t)_{u}\big), \; \hat{X}_{p}^{(s)}(t)_j = \underset{x}{\operatorname{argmin}}\sum_{u} \Omega^{s}_{ju} d\big(x,Z(t)_{u}\big),
\end{align*}  
with $1\leq s \leq S, 1\leq t \leq T$, and $\Omega^{s}$ is obtained from Step 2. In conjunction with data integration, inferring $\{Z(t), Y(t)\}$ in a low-dimensional latent space also allows us to identify and model the governing temporal pattern on conditional distribution of the response after integrating heterogeneous data. This strategy can be useful for other downstream tasks and applications, such as clustering and pattern exploration.  

\noindent \textbf{Step 4: Retrain the model with augmented training data} For the 
$s$th dataset at time $t$, we aggregate training data from all $S$ datasets reconstructed in the common latent space as $$\big(\tilde{X}^{(s)}_l(t),\tilde{Y}^{(s)}_l(t)\big) = \underset{1\leq s \leq S}{\cup}\big(\hat{X}_l^{(s)}(t), Y^{(s)}_l(t)\big),$$ 
and train the classifier $f(\cdot)$ on the augmented training dataset $\big(\tilde{X}^{(s)}_l(t),\tilde{Y}^{(s)}_l(t)\big)$ as $\tilde{f}_t^{(s)}(\cdot)$. Consequently, we update the prediction on unlabeled data $X^{(s)}_p(t)$ as $\hat{Y}^{(s)}_p(t) = \tilde{f}_t^{(s)}\big(\hat{X}^{(s)}_p(t)\big)$.     
}

\textbf{Remark}: our proposed method relaxes the 
assumption on the bijection mapping between joint distribution $(X,Y)$ cross-datasets. Instead we assume that there exist common clustering structures among distributions of covariates $X$ from different datasets. Specifically, for covariates $X^{(s)}$ and $X^{(s')}$ from the $s$th and $s'$th dataset, there exists a cross-dataset mapping $\Gamma: \bm{X}^{(s)}\to  \bm{X}^{(s')}$ which approximately preserves the metric information as 
\begin{align}\label{re_1}
d^{\left(s\right)}\left(X^{(s)}_i, X^{(s)}_j\right)\approx d^{(s')}\left(\Gamma\left(X^{(s)}_i\right), \Gamma\left(X^{(s)}_j\right)\right),
\end{align}
where $d^{(s)}(\cdot)$ and $d^{(s')}(\cdot)$ are the metrics on dataset $s$ and $s'$, respectively. Under the context of neuroscience applications where $\{X^{(s)}\}$ stand for neural activities in different subjects, condition (\ref{eq(1)}) is a reasonable assumption based on recent findings that the correlations among neurons generally lead to strong clustering organizations in neural activity patterns on the population level \citep{zhang2018review,carroll2009prediction}. These underlying common clustering structures among neural activities imply the existence of metric preserve $\Gamma(\cdot)$. On the other hand, the marginal distribution $\mathbb{P}^{(s)}(\Gamma(\bm{X}^{(s)}))$ can still be different from $\mathbb{P}^{(s')}(\bm{X}^{(s')})$ due to subjects' heterogeneity.

{In addition, the proposed method assumes that correspondence between response $Y$ and clusters of covariates $X$ can be preserved across different datasets. Specifically, we assume that the condition distributions of response $Y$ are invariant as
\begin{align}\label{re_2}
\mathbb{P}^{(s)}(\bm{Y}\mid \bm{X}^{(s)}) \approx \mathbb{P}^{(s')}(\bm{Y}\mid \Gamma(\bm{X}^{(s)})).
\end{align}
Notice that (\ref{re_2}) is the covariate shift assumption, which is generally made in most domain adaption problems \citep{courty2017joint}. Under the context of neuroscience applications, existing studies show that the clustering structures of neural activities generally serve as common neural codes carrying external sensory information and neural processing statuses information\citep{berry2020clustering, kleindienst2011activity, huang2016clustering}, which support the assumption (\ref{re_2}) for our real application. The detailed optimization procedure and hyperparameter tuning are summarized in the Section 8 of Supplemental Materials.}

\section{Simulation Study}

In this section, we investigate the performance of the proposed data integration approach and temporal integration method for several classification problems, and compare the performance to a baseline model that uses the data from individuals. Here we choose two classifiers for the numerical comparisons. The first is the quadratic discriminative analysis (QDA), and the second is the label propagation algorithm (LP); the former is a supervised learning classifier, whereas the latter is a semi-supervised learning method utilizing the structure information on covariates.

\subsection{Simulations of data integration for classification}

In this simulation setting, we consider six different datasets where the label of data points $Y$ in each dataset can belong to one of three classes. The data generation is as follows. For each class $Y = k, (k= 1,2,3)$, the two-dimensional covariates $X$ are sampled from the mixture of Gaussian distributions as $X \mid (Y = k)  \sim 0.5 \times N(\mu^{(1)}_k,I) + 0.5 \times N(\mu^{(2)}_k, I)$
where $I = \big((1,0)^T,(0,1)^T \big)$ is the covariance matrix, and $\mu^{(1)}_k, \mu^{(2)}_k$ denote the class-specific means. Specifically, we set
{\small    
\begin{align}\label{setting_1}
\mu^{(1)}_1 = (-a,a), \mu^{(2)}_1 = (a,a), 
\mu^{(1)}_2 = (2a,0), \mu^{(2)}_2 = (a,-a),
\mu^{(1)}_3 = (-a,-a), \mu^{(2)}_3 = (-2a,0),
\end{align}}
where $a$ represents the separability of $X$ among three classes. For each dataset, we randomly generate $N$ samples, where $Y$ follows a multinomial distribution with three classes of equal probability, and $X$ are sampled from the previous mixture Gaussian distributions. Then we obtain data points $(\Gamma_l X,Y)$ as the $l$th dataset, where $\Gamma_l (l = 1\cdots, 6)$ are the rotation matrices
$\Gamma_l = \big( (\text{cos $\theta_l$}, -\text{sin $\theta_l$})^T,(\text{sin $\theta_l$},\text{cos $\theta_l$}^T \big)$, and $\theta_l = 20\times(l-1)$. Notice that the covariates $X$ from each dataset have six clusters corresponding to different labels $Y$; we index these clusters as $c = 1,\cdots, 6$. Next we split data points from each dataset into a training set with $N_{train} = \sum_{s=1}^6 n_l^{(s)}$ data points, and a testing set with $N_{test} = \sum_{s=1}^6 n_p^{(s)}$ data points, where $N_{train} + N_{test} = N$. For the first dataset, we sample $90\%$ of the training data from clusters $c = \{1,3,6\}$ and $10\%$ from the other three clusters. Then the remaining data points serve as testing data. The generation of the training and testing data for other five datasets follows the same procedure, while the training data are sampled from different clusters. Specifically, the cluster indexes of training data for the second to sixth dataset are $\{1,4,6\},\{2,3,5\},\{2,3,6\}$,$\{1,4,5\},\{2,4,5\}$, respectively. The above data rotation and training data sampling scheme increase the heterogeneity of the distribution of $X$ across different datasets.    

In the following, we compare the classification performance of the data integration procedure in Section 3.1 with that of the individual learning scheme. For individual learning, we independently train a classifier on the training data, and then make predictions on the testing data from each dataset. While for the data integration scheme, we train the classifier for the $l$th dataset by using the training data from both the $l$th dataset and $M$ datasets. The number of datasets for integration varies as $M=1,3,5$. We then calculate the misclassification rate averaged from the 6 datasets using two different schemes. The comparisons are conducted under different sizes of training set $N_{train}$ and testing set $N_{test}$. In addition, we measure the separability intensity of the classification as pairwise distances among cluster centers as $\Delta: = \sum_{1\leq k_1 \leq k_2 \leq 3}\sum_{1\leq i \leq j \leq 2} \|\mu^{i}_{k_1} - \mu^{j}_{k_2}\|_2$, and denote the intensity as $\Delta_0$, $\Delta_1$, $\Delta_2$ when $a$ in (\ref{setting_1}) varies among $0.7,1,1.5$. A larger $a$ indicates a  strong separation of classifications.


\begin{table}[h]
\centering
\begin{adjustbox}{width=1
\textwidth}
\begin{tabular}{| c | c | c | c | c | c | c | c | c | c | } 
   \hline
   \hline 
\multirow{3}{*}{Signal intensity} & \multirow{3}{*}{$N_{train}/N_{test}$} &    \multicolumn{4}{|c|}{LP} & \multicolumn{4}{|c|}{QDA}  \\ \cline{3-10}  

& & Ind. & \multicolumn{3}{|c|}{Prop.} & Ind. &\multicolumn{3}{|c|}{Prop.} \\ \cline{3-10}

&   &   & $M = 1$  & $M=3$  & $M=5$ &  & $M=1$  & $M=3$ & $M=5$  \\  \cline{1-10}

\multirow{3}{*}{$\Delta_0$}  & 90/120  & 0.33(0.02)  & 0.30(0.02)  & 0.29(0.02) & 0.29(0.02)& 0.32(0.02) & 0.29(0.02)& 0.28(0.02) & 0.28(0.02)  \\ \cline{2-10}

&  150/150  & 0.33(0.02) & 0.30(0.02) & 0.29(0.02) & 0.28(0.02) &0.33(0.02) & 0.29(0.02) & 0.28(0.02) & 0.27(0.02)  \\ \cline{2-10}

& 240/180  & 0.36 (0.01) & 0.33(0.02) & 0.31(0.02) & 0.30(0.02)&
0.37(0.01) & 0.30(0.02)& 0.29(0.02) & 0.28(0.02) \\ \hline

\multirow{3}{*}{$\Delta_1 = 1.4{\Delta_0}$}  & 90/120  & 0.25(0.02)  & 0.20(0.02)  & 0.18(0.02) & 0.18(0.02)& 0.23(0.02) & 0.19(0.02)& 0.17(0.02) & 0.17(0.02)  \\ \cline{2-10}

&  150/150  & 0.26(0.01) & 0.20(0.02) & 0.18(0.02) & 0.18(0.02) &0.24(0.02) & 0.18(0.02) & 0.16(0.02) & 0.16(0.01)  \\ \cline{2-10}

& 240/180  & 0.32 (0.02) & 0.25(0.03) & 0.23(0.03) & 0.22(0.02)&
0.32(0.02) & 0.22 (0.02)& 0.21(0.02) & 0.20(0.02) \\ \hline

\multirow{3}{*}{${\Delta_2} = 2.2{\Delta_0}$}  & 90/120  & 0.17(0.02)  & 0.09(0.02)  & 0.07(0.02) & 0.06(0.02)& 0.14(0.03) & 0.07(0.02)& 0.07(0.02) & 0.07(0.01)  \\ \cline{2-10}

&  150/150  & 0.20(0.02) & 0.11(0.04) & 0.08(0.02) & 0.08(0.01) &0.16(0.02) & 0.07(0.01) & 0.07(0.02) & 0.06(0.01)  \\ \cline{2-10}

& 240/180  & 0.29 (0.01) & 0.21(0.06) & 0.19(0.05) & 0.17(0.05)&
0.29(0.02) & 0.14 (0.04)& 0.13(0.03) & 0.12(0.02) \\ \hline

\hline
   \hline
\end{tabular}
\end{adjustbox}
\caption{Misclassification rate over 30 repeated experiments from the individual learning scheme (Ind.) and data integration scheme (Prop.), where the data integration utilizes $M$ datasets. The misclassification rate reported in Table 1 is the average over 6 different datasets.}\label{4_1_1} 
\end{table}

Table \ref{4_1_1} shows that the proposed data integration method can substantially reduce the misclassification rate for both supervised learning classifier QDA and semi-supervised learning classifier LP by $9\%$ to $62\%$ over different simulation schemes. 
Notice that the improvement from the data integration increases as the number of incorporated datasets and the size of the training set within $N_{{train}}$ increase. This improvement pattern implies that the proposed alignment scheme captures the underlying common conditional distribution of label $Y$ given covariate $X$, which boosts the effective sample size of training data on each dataset. In addition, the improvement of our method is more significant when the separability of $X$ across different classes is strong, implying that the improvement also depends on the signal intensity of the shared response-covariate patterns within each individual dataset. In addition, we perform an additional simulation study where the labels are unbalanced in training sets. The results show that the proposed method achieves lower misclassification rate compared with the individual learning method. The detailed simulation setup and results are in the Section 12 of Supplemental Materials.

\subsection{Simulations for temporal integration for multiple datasets}

In this subsection, we consider simulation studies with the two-dimensional covariates dynamically changing over time. We compare the classification performance of integrating the temporal data procedure in Section 3.2 with that of the individual learning approach. For individual learning, we independently train a classifier on the training data, and then make a prediction on the testing data from each dataset at each time point. For our temporal integration scheme, we train the classifier on the common latent space integrating aligned training data from different datasets and time points. We then calculate the misclassification rate averaged from the six temporal datasets using these two approaches.  

Specifically, we consider data points with three classes $Y = k (k= 1,2,3)$. For each dataset at time point $t=1$, we generate the covariates in the same way as Section 4.1, and   
use $X^{(1)}$ to denote the observed $X$ at time $t=1$. To mimic the temporal pattern in $X$ over time points $t = 1,\cdots, T$, we propose a sequential update on $X^{(t)}$ based on $X^{(t-1)}$ as $X^{(t)} = \bm{b} + \bm{A}X^{(t-1)}$, where $\bm{b}$ introduces a global drift of $X$ from time point $t-1$ to $t$, and $\bm{A}$ provides a transformation on $X^{(t-1)}$. The stochastic drift term is $\bm{b} = (b_1, b_2)$, where $b_1, b_2$ are sampled from uniform distribution $\text{Unif}(0,0.4)$. The transformation matrix $\bm{A}$:
\begin{align*}
\bm{A} = \begin{pmatrix} 1 + \lambda & 0\\0 & 1 + \lambda\end{pmatrix} \begin{pmatrix} \text{cos} \theta & -\text{sin}\theta \\ \text{sin}\theta & \text{cos}\theta\end{pmatrix},
\end{align*}
where $\lambda$ and $\theta$ are randomly sampled from $\text{Unif}(-0.2,0.2)$ and $\text{Unif}(-0.2\pi,0.2\pi)$, respectively. Notice that the labels of the data points do not change during the process. Therefore, we construct a data sequence $(X^{(1)},X^{(2)},\cdots, X^{(T)},Y)$ for each dataset. Then we split data points from each dataset into $30$ training samples and $90$ testing samples following the same procedure discussed in Section 4.1. In
the following, we perform numerical experiments under different lengths of time $T = 4,8,12,20$, using different numbers of integrated datasets $M = 1,3,5$. 
\begin{table}[h]
\centering
\begin{adjustbox}{width=0.85\textwidth}
\begin{tabular}{| c | c | c | c | c | c |} 
   \hline
   \hline 
\multirow{3}{*}{Length of time} & \multirow{3}{*}{Num of datasets} &    \multicolumn{4}{|c|}{Classifier} \\ \cline{3-6}  

& & \multicolumn{2}{|c|}{LP} &\multicolumn{2}{|c|}{QDA} \\ \cline{3-6}

&  &  Ind. & Prop. & Ind. & Prop. \\ \hline

\multirow{3}{*}{$T = 4$}  & $M = 1$  & 0.146(0.02)  & 0.141(0.02)  & 0.156(0.01) & 0.179(0.01)   \\ \cline{2-6}

&  $M = 3$  & 0.176(0.01) & 0.133(0.01) & 0.184(0.01) & 0.139(0.01) \\ \cline{2-6}

& $M = 5$  & 0.169 (0.01) & 0.125(0.01) & 0.174(0.01) & 0.118(0.01) \\ \hline

\multirow{3}{*}{$T = 8$}  & $M = 1$  & 0.142(0.02)  & 0.145(0.02)  & 0.156(0.01) & 0.153(0.01)  \\ \cline{2-6}

&  $M = 3$  & 0.182(0.01) & 0.128(0.01) & 0.184(0.01) & 0.121(0.01) \\ \cline{2-6}

& $M = 5$  & 0.172 (0.02) & 0.121(0.01) & 0.174(0.01) & 0.113(0.01) \\ \hline

\multirow{3}{*}{$T = 12$}  & $M = 1$  & 0.142(0.01)  & 0.145(0.02)  & 0.156(0.01) & 0.137(0.01) \\ \cline{2-6}

&  $M = 3$  & 0.188(0.01) & 0.135(0.01) & 0.184(0.01) & 0.115(0.01)  \\ \cline{2-6}

& $M = 5$  & 0.176 (0.01) & 0.127(0.01) & 0.174(0.01) & 0.108(0.01)\\ \hline

\multirow{3}{*}{$T = 20$}  & $M = 1$  & 0.132(0.00)  & 0.064(0.00)  & 0.087(0.02) & 0.084(0.01) \\ \cline{2-6}

&  $M = 3$  & 0.161(0.00) & 0.108(0.01) & 0.117(0.02) & 0.093(0.01)  \\ \cline{2-6}

& $M = 5$  & 0.157 (0.00) & 0.104(0.01) & 0.183(0.01) & 0.095(0.01)\\ \hline

\hline
   \hline
\end{tabular}
\end{adjustbox}
\caption{Misclassification rates from the individual learning approach (Ind.) and the proposed temporal data integration scheme (Prop.) under different lengths of time $T$, where the proposed method utilizes $M$ datasets. QDA and LP stand for quadratic discriminative analysis and label propagation algorithm, respectively.}\label{4_2_1}
\end{table}

Table \ref{4_2_1} shows that the proposed temporal data integration method can reduce the misclassification rate for both classifiers QDA and LP compared with the individual learning even when the length of time is short $(T = 4)$ or the number of incorporated datasets is small $(M = 1)$. These results demonstrate the effectiveness of temporal alignment in reducing the systematic heterogeneity even with a small sample size. For both the QDA and LP classifiers, the improvement from temporal integration increases as $T$ increases given that $M$ is fixed. On the other hand, the improvement also increases as $M$ increases given that $T$ is fixed. 
In addition, when $T=4$, the classification improvement from our method over the individual learning method increases about $ 2.5\%$ when $M$ increases from 3 to 5. On the other hand, when $T=20$, the improvement increases about $13\%$ when $M$ increases from 3 to 5 on averaging two classifiers. This result implies that the proposed method can utilize common temporal patterns on the conditional distribution of response to further enhance the power of detecting the patterns shared across different datasets.     

{Notice that by applying the proposed method in Section 3.2, we obtain the sequential latent space to capture the common dynamic patterns of observed covariates and response-covariate association. To measure the performance of temporal alignment over all time windows, we investigate the misclassification rates of each time point, and calculate the average misclassification rates on the trajectory. A lower average misclassification rate indicates that the aligned temporal pattern can better capture the overall dynamic of response-covariate association.}

\subsection{Simulations for Synthetic Neuronal Spike Data}

{In this subsection, we consider the simulation experiments to mimic the application of the proposed data integration method to real neuronal spike data, where we simulate the neuronal spike data by following the widely used Poisson generalized linear modeling 
and Latent variable modeling \citep{keeley2020modeling}. 
Specifically, we consider total $M = 6$ datasets, where each consists of spike counts over $T = 20$ time points. The neuronal spikes of the $m$th dataset at the $t$th time 
point are governed by dynamic low-dimensional latent status $X_t^{(m)} \in R^{k},m =  1,\cdots, 6, t = 1,\cdots, 20$, where $k$ denotes the number of latent factors. The spike counts for the $j$th neuron at the $i$th trial are generated as follows:
$S_{j}^{(i)}(t) \mid X_t^{(m)} \sim \text{Poiss}\big(\sigma(W_{ij}X_t^{(m)} )\big), i = 1,\cdots, n, \; j =  1,\cdots, p$,
where $W_{ij}\in R^{p}$ denotes the loading vector, and $\sigma(\cdot) = \frac{\exp(\cdot)}{1+\exp(\cdot)}$ is the link function. Here we consider
the number of trials $n = 60$ and number of neurons $p = 60$. 
The latent status $\{X_t^{(m)}\}$ are sampled from mixture of Gaussian distributions with multiple clusters corresponding to different external stimuli. The detailed generation procedure for $\{X_t^{(m)}\}$ is the same to the setup in Section 4.2.} 

{We perform stimuli classification based on the labeled spike data $\{S^{(i)}(t), Y^{(i)}\}_{i=1}^n$. Specifically, we follow the proposed integrated latent alignment framework by first utilizing an autoencoder to achieve dimension reduction of the spike data $\{S^{(i)}(t)\}$ for each dataset in a sliding time interval with four time points. Then classifiers are trained on the integrated data points, and are compared with the individual learning method. We consider the proposed method integrating spike data from different numbers of datasets and time points, and Table \ref{new_sim} illustrates the average misclassification rates where Ind. denotes the individual learning approach, and Prop. 
denotes the proposed scheme. Table \ref{new_sim} shows that the proposed temporal data integration method can reduce the
misclassification errors for both classifiers QDA and LP compared with the independent training scheme. For both classifiers, the improvement from the proposed method increases as $T$ or $M$ increases, and the 
improvement maximizes at about $25\%$ when $T=16$ and $M = 6$. The simulation results demonstrate the effectiveness of the temporal alignment in reducing the systematic heterogeneity in subject-wise spike data.}

\begin{table}[H]
\centering
\begin{adjustbox}{width=0.85\textwidth}
\begin{tabular}{| c | c | c | c | c | c |}
   \hline
   \hline 
\multirow{3}{*}{Length of time} & \multirow{3}{*}{Num of datasets} &    \multicolumn{4}{|c|}{Classifier} \\ \cline{3-6}  

& & \multicolumn{2}{|c|}{LP} &\multicolumn{2}{|c|}{QDA} \\ \cline{3-6}

&  &  Ind. & Prop. & Ind. & Prop. \\ \hline

\multirow{3}{*}{$T = 4$}  & $M = 2$  & 0.403(0.02) & 0.413(0.02)   & 0.484(0.01) & 0.465(0.01)    \\ \cline{2-6}

&  $M = 4$  & 0.411(0.01) & 0.365(0.03)   & 0.463(0.01) & 0.435(0.02) \\ \cline{2-6}

& $M = 6$  & 0.459(0.01) & 0.408(0.05)   & 0.466(0.01) & 0.420(0.02) \\ \hline

\multirow{3}{*}{$T = 10$}  & $M = 2$  & 0.383(0.02) & 0.371(0.02)     & 0.401(0.01) & 0.381(0.03) \\ \cline{2-6}

&  $M = 4$  & 0.410(0.02) & 0.307(0.01)   & 0.414(0.01) & 0.325(0.03) \\ \cline{2-6}

& $M = 6$  & 0.437(0.02) & 0.319(0.03)   & 0.428(0.01) & 0.321(0.03) \\ \hline

\multirow{3}{*}{$T = 16$}  & $M = 2$  & 0.378(0.01) & 0.359(0.01)     & 0.376(0.03) & 0.346(0.03) \\ \cline{2-6}

&  $M = 4$  & 0.385(0.01) & 0.299(0.02)   & 0.392(0.01) & 0.283(0.03) \\ \cline{2-6}

& $M = 6$  & 0.411(0.01) & 0.310(0.02)  & 0.406(0.01)& 0.294(0.03) \\ \hline

\hline
   \hline
\end{tabular}
\end{adjustbox}
\caption{Misclassification rates from the individual learning approach (Ind.) and the proposed temporal data integration method (Prop.) under different number of time points $T$, where the proposed method integrates $M$ datasets. Here QDA and LP stand for quadratic discriminative analysis and label propagation algorithm, respectively. The spike data are simulated based on Poisson generalized linear modeling and latent variable modeling.}
\label{new_sim}  
\end{table}

\section{Real Data Application}

In this section, we apply the proposed ILA framework to the rodent electrophysiological study introduced in Section 2. Here we use spike data from five rodents recorded as the rodents  smelled different odors in the sequence memory task. The spike data from the $l$th rat is reformulated as a three-dimensional data array $\mathcal{A}^{(l)} = \{A_{ijk}\}, l = 1,2,3,4,5$ where the $i,j,k$ represent the trials, neurons, and time points for each rodent, respectively. The spike data $\mathcal{A}^{(l)}$ consists of binary elements where $A_{ijk} = 1$ indicates that the specific neuron fired, and $A_{ijk} = 0$ that it did not fire. In each trial, the subject samples one of five odors labeled "A", "B", "C", "D", "E". We then extract 4 seconds of spike data per trial (2 s before and after odor onset) in bins of 10 ms, resulting in 400 time points for each trial. In the following, we investigate a subset of the original spike data from trials corresponding to the four labels "A", "B", "C", "D" (odor E was excluded because of its lower number of trials) and time interval between the $250$ millisecond and $500$ millisecond after the onset of odors. This time interval of the data was shown to exhibit the strongest differentiation of odor information, as well as significant differentiation of neuronal coding for different temporal order and trial outcome information (\cite{shahbaba2022hippocampal}). In summary, the dimensions corresponding to trials, neurons, and time points of spike data for the five rodents are $168 \times 46 \times 25$, $152 \times 49 \times 25$, $207 \times 104 \times 25$, $133 \times 92 \times 25$, and $174 \times 79 \times 25$, respectively. 

In the following, we perform the first stage in the ILA framework to compress the original temporal spike data into a low-dimensional latent space for each rodent by using autoencoder, which is 
a nonlinear dimensionality reduction method based on neural networks. 
The implementation details of the autoencoder are in the Section 10 of the Supplemental Materials.  
Through the above process, the spike data from each $100$ms sub-window are compressed into two-dimensional latent features. We collect the preprocessed spike data from the 16 sub-windows for the $s$th rodent $(s = 1,2,3,4,5)$ as $(X^{(s)}(1),X^{(s)}(2),\cdots,X^{(s)}(16),Y^{(s)})$, where $X^{(s)}(t) \in R^{n_l \times 2}$ denotes the latent neural representation of the $s$th rat at the $t$th sub-windows $(t = 1,\cdots, 16)$ and $n_l$ is the number of trials from the $l$th rat. In addition, $Y^{(l)} \in \{1,2,3,4\}^{n_l}$ denotes the odor labels "A", "B", "C", "D" corresponding to each trial of the $l$th rat. In the following, we first apply the proposed integrated supervised learning method in Section 3.1 to the preprocessed spike data, and compare the results with that of individual learning on the odor classification. Specifically, we sample $20\%$ of the trials from each rat as training data, and the remaining $80\%$ of trials serve as testing data. Then we train a classifier by our integration method and the individual learning method separately, and investigate the corresponding misclassification rates for the two methods on each sub-window. We use the K-nearest neighbors algorithm (KNN) as our classifier following the suggestion in \citep{shahbaba2022hippocampal} due to the nonlinear boundary among the latent neural representations from different odors. {The number of neighbors for KNN is set as 25 labeled data points for both the individual learning method and our method; i.e., the label of a specific data point is determined by the labels of the 25 nearest data points, where the number of neighbourhoods follows the setup in \citep{shahbaba2022hippocampal}. 



Figure \ref{fig_3_acc_2} illustrates the classification accuracy of KNN trained at each $100ms$ sub-windows by the individual learning method (blue line) and the proposed integration learning method (red line), respectively. The odor classification analysis is performed for each 100-ms time window based on the output from the data compression stage, i.e., the autoencoder. Specifically, the autoencoder compresses the original neural spike data within each 100-ms time window into two-dimensional latent features, and the 100-ms time window slides from 250 ms to 500 ms with a step of 10 ms. The results show that the proposed method achieves higher odor classification accuracy for all five rodents over the sequential sub-windows. 
The improvement from the proposed data integration suggests the neuronal ensembles from the five rodents have in common the fact that they exhibit a differentiation of odor information in that time window. More specifically, it shows that our approach can detect the common neuronal coding among the different subjects, and utilizes the shared pattern to enhance the performance of odor classification. In addition, the results show significant improvements from the proposed method for rat 1, rat 4, and rat 5 across all time intervals. For rat 2 and rat 3, the improvements are more significant in the second half of the time intervals, while improvements in the first half of the time intervals are modest. We also investigate the improvement of the average classification accuracy among the five rats, which is illustrated in the bottom right of Figure \ref{fig_3_acc_2}. The result shows that the data integration method achieves about $10\%$ higher average classification accuracy compared with the individual learning method. For our specific application, this is a substantial improvement.


\begin{figure}[h]
\begin{center}
\includegraphics[width = 5.6in, height = 2.8in]{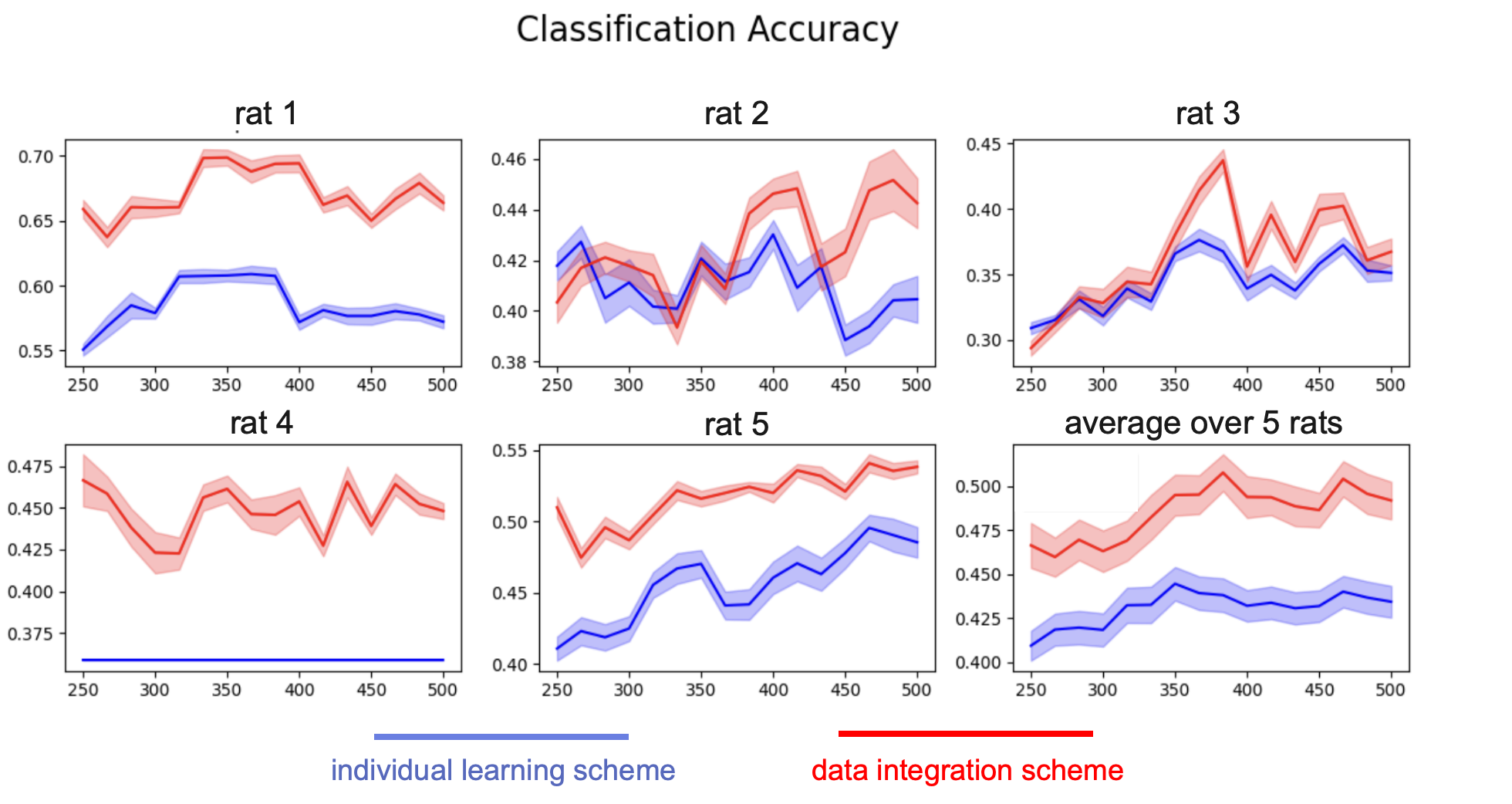}
\end{center}
\caption{{\small The four-odor classification accuracy based on the three rodents' latent neural features extracted by autoencoder within time interval $250\sim 500$ milliseconds. The blue line denotes the classification accuracy from KNN trained by the individual learning method at each time point. The red line denotes KNN's classification accuracy trained by the proposed data integration method in Section 3.1.}}

\label{fig_3_acc_2}
\end{figure}

In the following, we investigate the odor classification performance of the temporal integration method introduced in Section 3.2, and compare the performance of the individual learning approach based on the \textbf{five} rodents spike samples within the time interval $250\sim 500$ milliseconds. For each rodent, we sample the odor labels and the corresponding latent neuronal coding from $20\%$ of the trials as training data, and the remaining $80\%$ of the trials as testing data. 
Then we train the KNN classifier by our temporal integration method and individual learning method separately, and examine the misclassification rates of the two methods on each 16 sub-windows. 
In addition, the number of template samples in the aligned space is set as $n=200$ for the temporal alignment method. 

We use the individual learning scheme as the baseline, and the classification comparison is illustrated in Figure \ref{fig_4_acc}. The results show about $10\%$ accuracy improvements from the proposed method for rodent 2 and rodent 4 across all time intervals. For rodent 1, rodent 3, and rodent 5, the accuracy of the proposed method is higher than the individual learning method at most of the time points, and the improvements can be about $10\%$ at some time windows. In addition, compared with the individual learning method, the accuracy of our method changes more smoothly over the time because the proposed method utilizes the time dependency among latent representations of spike samples. This is another advantage of the proposed method in scientific interpretation in addition to improving classification accuracy. Notice that the non-temporal data alignment method introduced in Section 3.1 is designed for response classification at each time point, while the temporal alignment method introduced in Section 3.2 aims to capture the common dynamic patterns of covariate-response association better and more smoothly. In other words, the temporal alignment method in Section 3.2 does not necessarily achieve better timepoint-wise response classification than the non-temporal method, and vice versa.

\begin{figure}[h]
\begin{center}
\includegraphics[width = 5.6in, height = 2.8in]{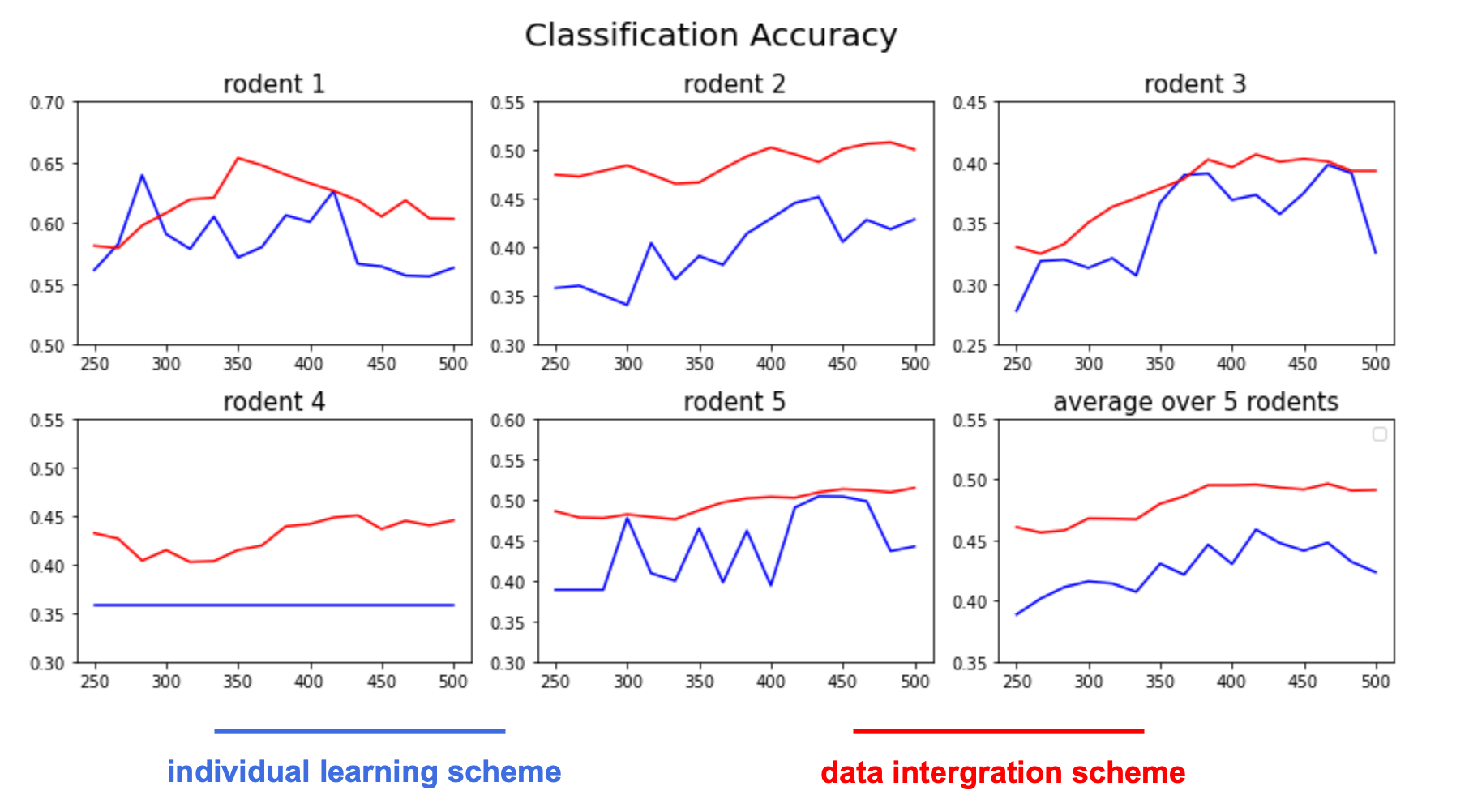}
\end{center}
\caption{{\small The four-odor classification accuracy within time interval 
$250 \sim 500$ milliseconds based on rodents' electrophysiological data. The blue line denotes the classification accuracy from KNN trained by the individual learning method at each time point. The red line denotes KNN's classification accuracy trained by the proposed temporal integration method in Section 3.2.}}\label{fig_4_acc}
\end{figure}

Notice that Figure \ref{fig_3_acc_2} and \ref{fig_4_acc} correspond to the classification-oriented alignment method presented in Section 3.1, and the covariate-response association dynamic alignment method presented in Section 3.2. The non-temporal data alignment method introduced in Section 3.1 is designed for response classification at each time point, while the temporal alignment method introduced in Section 3.2 aims to capture the common dynamic patterns of covariate-response association better and more smoothly. In other words, the temporal alignment method in Section 3.2 does not necessarily achieve better timepoint-wise response classification than the non-temporal method and vice versa. Therefore, instead of comparing alignment methods in Section 3.1 and Section 3.2, Figure \ref{fig_3_acc_2} and Figure 4 compare each of the proposed methods with the individual classification scheme, and demonstrate that both of the proposed methods can improve task-relevant response classification over the individual method. In contrast, the advantage of temporal alignment method in Section 3.2 is to allow one to investigate the underlying common covariate-response association that smoothly changes over time, which is illustrated in following Figure \ref{Trajectories_1}.

In summary, the proposed alignment approach can integrate information from both different subjects and different time windows. The improvement from the temporal data integration on the five subjects implies the existence of a common structure in the neuronal coding associated with the processing of the different odors. More importantly, the dynamic of the neuronal coding clustering is also shared across subjects to some extent. This finding is consistent with prior evidence that hippocampal neurons represent information about task events critical for accurate performance \citep{allen2013evolution, eichenbaum2014time}.


We also utilize the temporal data integration in Section 3.2 to investigate the common dynamic pattern of neuronal coding among the five rodents. Specifically, we first obtain the rodent-wise alignment mapping $\{\Omega^{(s)}\}_{s=1}^5$ and temporal similarity matrices $\{D(t)\}_{t=1}^{16}$ of 16 sub-windows from Stage 2 in the temporal integration where we use the true odor labels to construct the label distance matrices $\{M^{(s)}(t)\}, s = 1,\cdots,5,\; t = 1,\cdots, 16$. Then we reconstruct the aligned latent space of neuronal coding for the five rodent at the $250 \sim 500$ms time windows by following Stage 3 based on $D(t)$ and $\Omega^{(s)}$. Finally, we fix the estimated cross-subject latent space from the time window at $250$ to $500$ms, and then project neuron spike samples from the other time windows throughout $0$ to $1500$ms into this latent space. Figure \ref{Trajectories_1} shows neural activity trajectories for all 5 rodents throughout the $0$ to $1500$ms period. Notice that we investigated the odor misclassification rates by using common latent spaces at different dimensions $k=2,3,4$, and found that different dimensions provide very close rates. For the purpose of visualization, we adopt $k=2$ to illustrate the common dynamic patterns of the latent neural coding.

\begin{figure}[h]
\begin{center}
\includegraphics[width=0.9\textwidth]{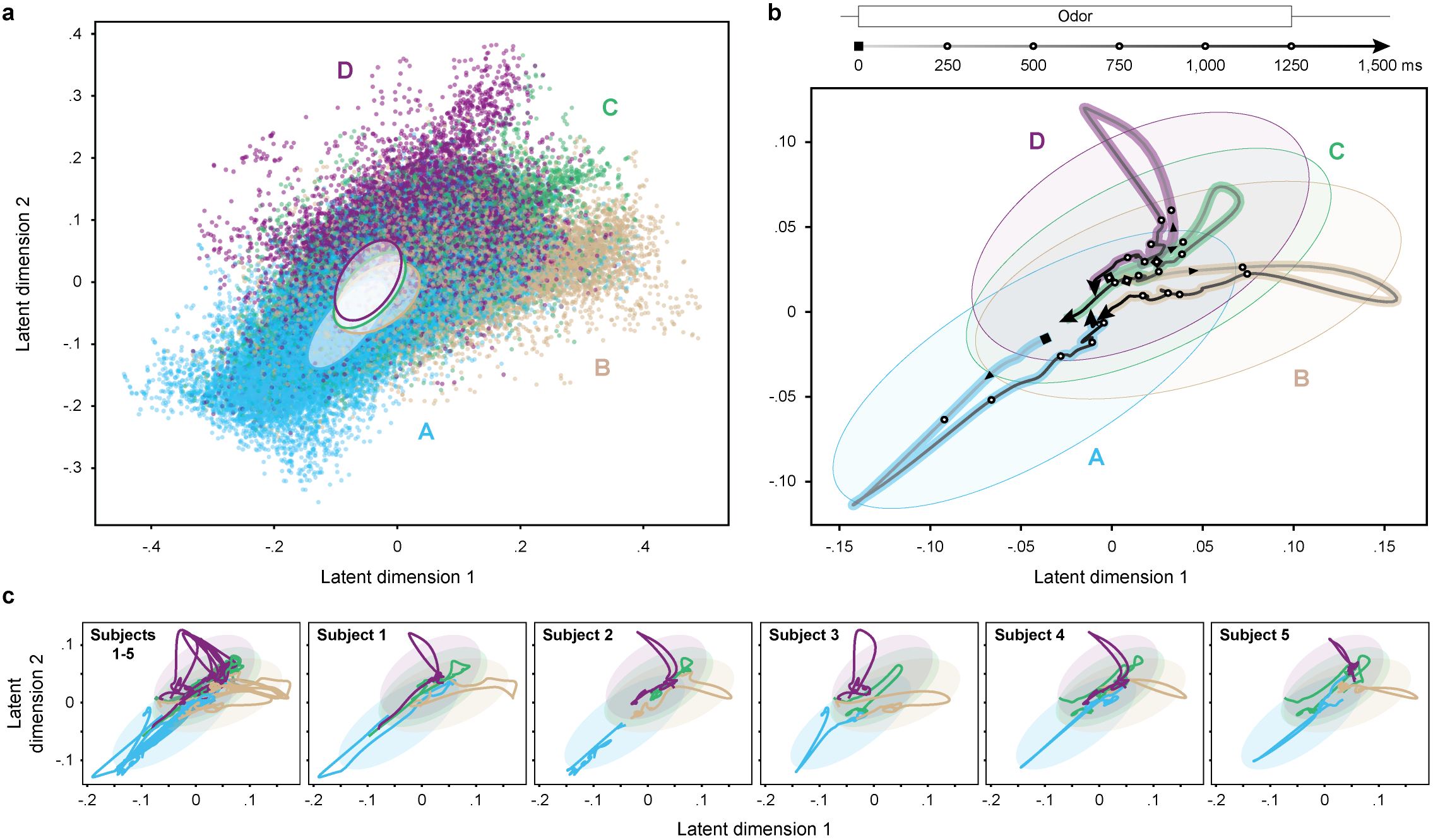}
\end{center}
\caption{{\small Odor-specific neuronal coding and dynamics in the aligned latent space.
 \textbf{a.} Differentiation of the neural activity across odor presentations on the aligned latent dimensions (odors A-D; data aggregated across all 5 subjects). Each dot indicates the location of the neural activity for a 100-ms bin from a given subject, with the color denoting the odor label of the corresponding trial.
 \textbf{b.} Neural activity trajectory (mean across subjects) on the latent dimensions during each odor trial type. 
 \textbf{c.} Neural activity trajectories for each subject showing comparable patterns across subjects. Ellipses are consistent across panels and represent the spread of the latent coding for each odor type, specifically the two largest square-rooted singular values from the covariance of latent coding within each cluster.}}
\label{Trajectories_1}
\end{figure}


{The results show that the neural activity on odor A trials is very distinct than that on odors B, C, and D trials. This can be explained by the fact that the sequence always starts with odor A (so animals are not required to remember it), and it is preceded by a running period (so that activity dynamics associated with running extend into the odor period). Therefore, we focus on comparing the coding properties and dynamics observed during odors B-D, in which the behavioral and cognitive demands are matched.}

{Neural trajectories have been reported in the motor system in the past \cite{linden2022movement}. In fact, the motor system was an ideal system to initially demonstrate such trajectories (or proof of concept) because of its simplicity due to a high correspondence between motor system  and direction of movement.}
However, capturing neural trajectories for cognitive information is this paper, 
is considerably more difficult -- neural activity in the hippocampus is several synapses (connections) away from the external world, and thus represents information that is more abstract, multi-dimensional and variable across subjects. 

{Notice that existing literature has utilized dimension-reduction method to visualize trajectory representing the response over time to each odor presentation \citep{brown2005encoding}. However, our data integration approach allows, for the first time, integration of such trajectories from different subjects into a common (aligned) latent space (Fig \ref{Trajectories_1}b). 

Specifically, the activity on odors B, C, and D trials starts at a similar location in the latent space and heads toward the center of their respective clusters (0-250 ms period), reaching maximum differentiation in the 250-500 ms period. Toward the end of the odor presentations (750-1,250 ms period), the activity for each odor then converges back toward the beginning of their trajectory (faster for odor D, than C, than B), before reaching a distinct region of the latent space in the post-trial activity (1,500ms; indicated by arrowheads). Our data integration approach also allows the visualization of each subject’s neural activity in the aligned latent space (Fig \ref{Trajectories_1}c), which reveals comparable trajectories across the 5 subjects. Critically, these results would have been difficult, if not impossible, to interpret without this data integration approach, as the neural activity of each subject would otherwise be in distinct latent spaces that could not be directly compared across subjects.}

\section{Discussion}

In this paper, we propose a novel and general data integration framework based on optimal transport for supervised learning tasks, which is motivated by a rodent electrophysiological experiment investigating non-spatial memory mechanisms. 
Our method is effective when the data domains and dimensionalities of each dataset are different, and the integration is effective even if the number of datasets is small. 

We apply the proposed framework to electrophysiological data collected from a rodent non-spatial memory study. The proposed method identifies the existence of common neuronal coding patterns across subjects during the processing of different odors. 
This discovery is consistent with previous findings that hippocampal neurons represent information about task events critical for accurate performance \citep{allen2013evolution, eichenbaum2014time}. Furthermore, our results suggest that heterogeneous neuron datasets can be systematically integrated to increase the discriminating power of an electrophysiological experiment, which may have broad implications in neuroscience research and beyond. 

\bibliographystyle{apa}
\bibliography{OT_reference,response, response_MR}

\end{document}